\documentclass{aastex}
\usepackage{spr-astr-addons}
\RequirePackage{color}

\def\msun{\ensuremath{M_{\odot}}}

\begin{document}

\title{On the distribution of initial masses of stellar clusters inferred from synthesis models}
\shorttitle{Initial mass cluster distribution function}
\shortauthors{Cervi\~no and Valls--Gabaud}
\author{M. Cervi\~{n}o} 
\affil{Instituto de Astrof\'{\i}sica de Andaluc\'{\i}a (CSIC), Apdo. 3004, Granada 18080, Spain}
\and 
\author{D. Valls--Gabaud}
\affil{GEPI, CNRS UMR 8111, Observatoire de Paris, 5 Place Jules Janssen, 92195 Meudon Cedex, France}

\begin{abstract}
The fundamental properties of stellar clusters, such as the age
or the total initial mass in stars, are often inferred from 
population synthesis models. The predicted properties are then
used to constrain the physical mechanisms involved in the
formation of such clusters in a variety of environments. Population
synthesis models cannot, however, be applied blindy to such
systems.  We show that synthesis models cannot be used in
the usual straightforward way to small-mass clusters (say, $M < $
few times 10$^4$ \msun). The reason is that the basic hypothesis 
underlying population synthesis (a fixed proportionality between the 
number of stars in the different evolutionary phases) is not fulfilled 
in these clusters due to their small number of stars. 
This incomplete sampling of the stellar mass function results in 
a non-gaussian distribution of the mass-luminosity ratio for 
clusters that share the same evolutionary conditions (age, 
metallicity and  initial stellar  mass distribution function).  
We review some tests that can be carried out a priori to check
whether a given cluster can be analysed with the fully-sampled
standard population synthesis models, or, on the contrary, 
a probabilistic framework must be used. This leads to a
re-assessment in the estimation of the  
low-mass tail in the  distribution function of initial masses
of stellar clusters. 
\end{abstract}

\section{Introduction}
\label{sec:intro}

The study of astrophysical objects is often limited by our ability 
to infer their physical properties such as distances, masses or ages, 
from their observed fluxes. In  stellar astrophysics, when
the distance to an observed star is known, the luminosity at different 
wavelengths or spectral energy distribution (SED), $l_\lambda$, provides 
constraints on the effective 
temperature  and on its mass-luminosity ratio. This ratio  
 depends mainly on the effective temperature,  gravity and the
mass of the star. In the case of  stellar clusters with known 
distances, the observed luminosity is the sum of the 
luminosities of the stars in the cluster, each one with its own mass-luminosity 
ratio. In a cluster composed by $N_{\star}$ stars, this luminosity can be written as 
\begin{equation}
L_\mathrm{cluster} = \sum\limits_{i=1}^{N_{\star}} l_i \quad .
\label{eq:Lindv}
\end{equation}
As stated in Eq~(\ref{eq:Lindv}), the integrated luminosity of the cluster does not 
provide a great deal of information on the stars in the cluster. However, we 
know that the possible luminosities and spectral shapes  of individual 
stars are in the range defined by stellar evolution, and thus it is 
possible to group individual stars in representative classes or 
evolutionary phases $j$ of luminosity $\ell_j$. Assuming a total number 
of $N_\mathrm{class}$ 
stellar evolutionary phases, Eq. \ref{eq:Lindv} can be rewritten as 
\begin{eqnarray}
L_\mathrm{cluster} &=& \sum\limits_{j=1}^{N_{\mathrm{class}}} n_j \, \ell_j \quad , \\
N_{\star} &=& \sum_{j=1}^{N_{\mathrm{class}}} n_j \quad .
\label{eq:Lclass}
\end{eqnarray}
The problem now becomes the estimation of the $n_j$ coefficients 
in such a way that we can obtain physical properties of the cluster from 
them. At this stage, it is not possible to know the total number of stars 
in the cluster from its integrated light, nor how many stars are in 
a given evolutionary phase. However, we can  relate the {\sl relative} number 
of stars in different evolutionary phases thanks to stellar evolution: 
 the number of stars in a given evolutionary phase is 
proportional to the amount of fuel that can be consumed in that phase, 
and therefore with the lifetime $t_j$ of the stellar evolutionary phase.   
This is the {\it Fuel Consumption Theorem} \citep{TG76,RB86,MG01} which underlies,
implicitely or explicitely, any population synthesis method. 

The comparison 
of different evolutionary phases (say, phase $i$ vs phase $j$) 
provides the number or population ratio $n_i/n_j$ in the 
limit of an infinite number of stars in the cluster. Indeed,  if the cluster actually 
has a very large number of stars in {\it all} the theoretical evolutionary phases,
we have that 
\begin{eqnarray}
\frac{t_i}{t_j} & \propto & \frac{w_i}{w_j} \quad , \\
w_k & = & \lim_{N_{\star}\rightarrow \infty} \left(\frac{n_k}{N_{\star}}\right) \quad . 
\end{eqnarray}
These relations hold for the  post main sequence evolutionary phases,
that is, the populations of the most luminous stars which dominate 
 the total integrated luminosity. Therefore once we explicit the relation between 
the $w_i$ coefficients and the most luminous stars, 
 the values of luminosity ratios ({\sl i.e.} colours) are also fixed.  
Since the proportionality relations between the stellar evolutionary 
phases that {\it would be} present in a cluster depend on the age of 
the cluster, in general $w_i = w_i(t)$. This provides a way to estimate the age of 
the given cluster, for example by the comparison of different colours 
of the cluster with theoretical predictions.

Note, however, that these relations do not allow us to obtain the total 
mass of the cluster unless 
the most luminous stars are also the ones that define the total mass
 ({\sl i.e.} the unlikely case where the most massive stars 
are also the most numerous ones, a hypothesis that is ruled out 
by the observations). So what is the fraction of 
the total mass which is responsible for the total luminosity? Since we 
do not know the actual stellar mass distribution function of the cluster, 
we have to use a statistical method to describe how many stars with 
a given initial mass are expected in the cluster: the stellar initial
mass function. For simplicity, 
 we assume that all the stars in the  cluster have been formed in
a single star formation episode, and that there are no other
episodes so that we do have a  single stellar 
population (SSP).  The integration 
of the stellar initial mass function over the mass range of initial masses 
that defines an evolutionary phase $i$, $m_i \pm dm_i$, provides the $w_i$ 
coefficients that allow us to obtain the mass-luminosity ratio of stellar 
clusters as a function of age as
\begin{equation}
\left(\frac{M}{L}\right) = \frac{\sum\limits_{i=1}^{N_{\mathrm{class}}} 
w_i \, m_i}{\sum\limits_{i=1}^{N_{\mathrm{class}}} w_i \, \ell_i} = 
\lim_{N_{\star}\rightarrow \infty} \frac{	\sum\limits_{i=1}^{N_{\mathrm{class}}} 
n_i \, m_i}{\sum\limits_{i=1}^{N_{\mathrm{class}}} n_i \, \ell_i} \quad .
\label{eq:ML}
\end{equation}
The inferred cluster initial mass is then obtained from this implicit
 mass-luminosity relation through a direct combination of the theoretical 
mean luminosity
\begin{equation}
L_\mathrm{theo} = \sum\limits_{i=1}^{N_{\mathrm{class}}} w_i \, \ell_i \quad ,
\end{equation}
\noindent with the inferred mass-luminosity ratio :
\begin{equation}
M_\mathrm{inferred}\;  = \; L_\mathrm{theo} \; \times \; \left(\frac{M}{L}\right) \quad .
\end{equation}
Strictly speaking, this direct comparison
 provides in fact the {\sl expected} number of stars in the cluster $<N>$, and the 
{\sl expected} mass $<M>$ which corresponds to this expected number of stars. 
It is important to note that an incorrect age estimation also implies
 an incorrect cluster mass estimation.

Synthesis models provide this mass-luminosity relation for different
 ages and metallicities\footnote{Note that this metallicity refers 
to the evolutionary tracks and not the  metallicity in the stellar atmospheres, 
which may not be the same.} 
 for theoretical clusters which contain an infinite number of  stars. 
This deterministic method has been used, rather blindly, to clusters
 of any mass, and in this case the mass-luminosity relation is a simple 
function of the age and metallicity ($M/L = f(t,Z)$), and the fundamental
properties of clusters 
 are inferred by statistical tests such as $\chi^2$ fits. 
This result is also recovered 
as the mean value of the distribution of the possible mass-luminosities 
relations, ($M/L = f(t,Z,M))$, under a probabilistic framework 
\citep{CL06}. In this case, it is necessary to take into 
account the shape of the distribution and how it varies with the 
cluster mass when synthesis models results are applied to
 the analysis of real clusters.

\section{The distributed of the mass-luminosity ratio in stellar 
clusters and the initial cluster mass function}
\label{sec:ML}

\begin{figure}[t]
\centering
\includegraphics[width=\columnwidth]{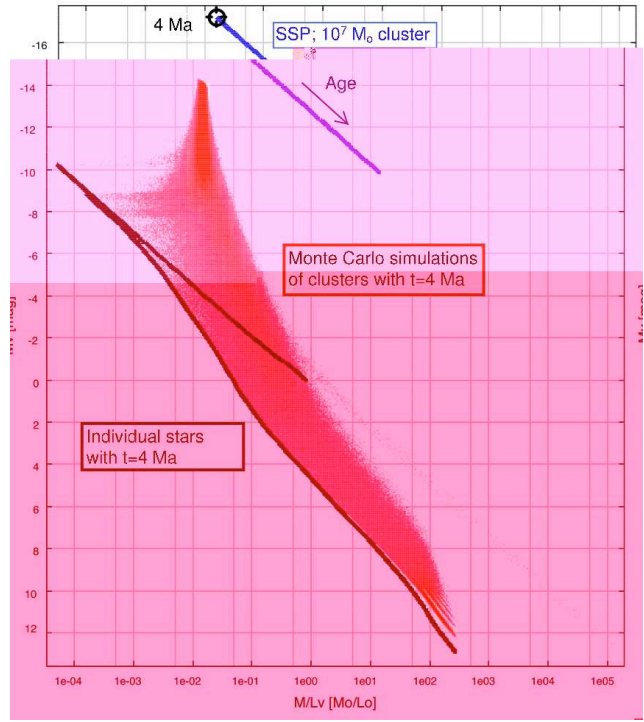}
\caption{Mass-luminosity relation in three sets of observables : (1) Top
thick blue line : a 10$^7$ M$_\odot$ 4 Ma-old SSP model (circle) which
slowly and monotonically evolves with age; (2) Bottom thick black line :
single stars in a 4 Ma-old isochrone, the upper branch being the post-main-sequence
stars, and the lower branch the main sequence; (3) Shaded area: 10$^6$ simulations of a population
of coeval  4 Ma-old clusters, whose initial mass function is a power-law
with slope --1, with lower and upper mass limits of 0.1 and 10$^5$ $M_\odot$
respectively. Note at the bottom  the sequences formed by binaries, triple
and quadruple systems, etc. Clearly, the behaviour of the mass-luminosity
ratio in this regime of small-mass clusters cannot be extrapolated
from the evolution of the more massive cluster at the top. }
\label{fig:fig1}
\end{figure}

The deterministic method is only valid, as seen above, in the
limit of a very large number of stars populating most, if not all,
the evolutionary phases. The blind application of the method
to small clusters results in wrong inferences have been made because the
underlying assumptions are violated. A simple illustration
is provided by  Fig. \ref{fig:fig1}, which 
 shows the mass-luminosity ratio for different cases as a 
function of the cluster luminosity in the $V$ band. The thick blue 
line at the  top shows the evolution with  age of the mass-luminosity ratio 
and the V magnitude for a 10$^7$ M$_\odot$ cluster obtained from the  
SSP models provided by  \cite{Gietal02}. The circle at the top of the 
line shows the position of a 4 Ma-old cluster and the monotonic decrease in
luminosity yields a monotonic increase in the mass-luminosity ratio,
a property often used to infer fundamental properties as described
above.
 The thick black line at the bottom 
of the plot shows the position of 4 Ma-old individual stars from the 
corresponding isochrone provided by  \cite{Gietal02}. Note that, because
of the use of the $V$ band, the turn-off point appears to be
the brightest with the smaller mass-luminosity ratio. The upper branch
are the post-MS stars, while the lower branch provides the locus
of the MS stars, down to very low luminosities and hence large mass-luminosity
ratios.
 The shadow region 
in the middle of the plot is the result of 10$^6$ Monte Carlo simulations 
for 4 Ma-old clusters using the same stellar initial mass function than 
the SSP models. The mass of each cluster is the result of a random 
sampling of a power-law initial cluster mass function with slope $\alpha = -1$ 
covering the cluster mass range between 0.1 and 10$^5$ M$_\odot$.  
The stellar initial mass function has been sampled randomly until 
the cluster mass of the cluster has been reached. 

Let us consider the mass-luminosity relation of individual stars of a 
given age (bottom thick black line in Fig. \ref{fig:fig1}, isochrone
of a 4 Ma-old population of single stars). Obviously, the possible 
values of the mass-luminosity depends on the particular properties of 
each star (its age, metallicity and mass; $M/L = f(t,Z,M)$). The figure 
shows that the range covered by the mass-luminosity relation of individual 
stars include all the possible mass-luminosity relations of SSP models. 
It also shows that the stellar mass-luminosity relation defines a 
natural limit of the mass-luminosity relations obtained in the Monte 
Carlo simulations of clusters.

When stars are combined to describe stellar clusters (following a given 
stellar initial mass function), the mass-luminosity relation {\it gradually}
 collapses to a {\sl single} mass-luminosity ratio. The origin of this 
evolution in the distribution function  of the mass-luminosity ratio
is simply  explained by  the 
right-hand side of Eq.~(\ref{eq:ML}): the actual fraction of  stars in 
a given evolutionary phase $n_i/N_{\star}$ does not coincide with the 
theoretical value $w_i$, but  fluctuates around it following a 
multinomial distribution (see \citet{CL06} for a detailed 
discussion, and \citet{CVG03} for a quasi-Poisson formalism). If $N_{\star} w_i$ 
is the expected number of stars in the $i$-th evolutionary phase,
 the $n_i$ value of real clusters will be distributed around it, 
producing  variations with respect to the expected total luminosity of 
the cluster, but almost no variation in its total mass. Equivalently, 
variations in the number of low-mass stars yield variations in the 
total mass, but not in the total luminosity. Obviously, the dispersion 
in the mass-luminosity ratio will be larger for clusters which have a smaller  
number of stars since these clusters have large relative dispersions  
in  $n_i $\citep{CLC00,CVGLMH02,CVG03,CL06}. Note that the distributed nature 
of the mass-luminosity relation is a result of the intimate composition
 of stellar populations of real stellar clusters. Its {\it physical}
 nature implies that it remains a distribution even in the case of perfect 
observations performed in perfect telescopes with perfect instruments
 with no statistical observational errors.

Only when the number of stars in a cluster is large enough ({\sl i.e.} the 
cluster is bright enough) the mass-luminosity ratio obtained by
 SSP models becomes a reliable, unique and well-behaved quantity. In other
 terms, the assumption of a mass-luminosity relation independent of the 
cluster mass is only valid for massive  clusters, typically with 
masses  larger than 10$^5$ $M_{\odot}$.

We want to stress that the main issue due to the incomplete 
initial stellar mass function sampling 
 is that the proportionality between the actual evolutionary 
phases in the cluster at a given age $t_1$, $n_i(t_1)/n_j(t_1)$,  differs from the  assumed
one in the synthesis models, $w_i(t_1)/w_j(t_1)$. Not only it may well be 
not fitted by the models, but it could also be close to the  
proportion $w_i(t_2)/w_j(t_2)$ that corresponds to a different 
age $t_2$.  For example, young clusters without massive stars 
(due to the sampling of the stellar initial mass function) 
 are systematically best fit by models at older ages 
because older clusters do not have massive stars. Under the usual
assumption of full sampling, the sparse sampling of the IMF in
these clusters is wrongly interpreted as a pure evolutionary effect.
As  the mass-luminosity 
ratio decreases with age, this effet translates into  an overestimation of 
the initial cluster mass, producing a systematic bias in the cluster mass estimation.

From another perspective,  when sampling effects are present, there is more
 information on the properties of {\it particular} stars in the clusters 
(the effective temperatures and luminosities of individual stars) but 
there is less information about the {\it global} properties of the system 
(age and cluster masses). We refer to \citet{Buzz93} and \citet{Buzz05} 
for a more detailed analysis on the information that can be obtained
 from a stellar population  through synthesis models.

In the case of extreme sampling effects, the integrated light does not 
provide any information about the cluster, and accurate age or mass 
determinations can only be done taking into account the theoretical probability 
distribution functions that will produce a distribution of possible physical
 properties compatible with the observations. The range of physical 
properties will be larger when the number of stars in the cluster is 
smaller (the range of stellar mass-luminosity ratio is larger than 
the range predicted by  SSP models), and implies an {\it intrinsic loss of precision} in the
 global properties of the cluster \cite[see ][ for a more extended
 discussion]{CL07}. The only way to estimate precise ages in this
 situation is to obtain the most detailed information about the number
 of stars in each evolutionary phase, that is, to analyse the 
colour-magnitude diagram ({\sl i.e.} the individual stars) of the clusters
 \cite[{e.g.}][]{Peletal06,HVG08}. Unfortunately, the colour-magnitude analysis 
is not reliable for obtaining cluster masses, which are controled by low-luminosity stars.

An alternative choice for a rough estimation of cluster ages is to 
look for signatures that are only present in a limited temporal range. 
In the case of young clusters, an example would be to look for young star 
signatures, such as Wolf-Rayet features or emission lines: the presence 
of these signatures implies a young cluster, but the absence of these
 signatures does not imply an old cluster, just the absence of massive
 stars ! Again, this rough age estimation does not provide information
 about the cluster mass.

\subsection{A simple test for  sampling effects identification}

The most trivial test to identify when sampling effects are essential 
 for an accurate analysis is to use  the Lowest Luminosity Limit (LLL)
method described in  \citet{CL04}. The LLL implies that it has no meaning 
to compare a  cluster with synthesis models (in a deterministic way)
 if the integrated luminosity of the cluster is lower than the luminosity 
of the most luminous star included in the model. This simple statement 
restricts the deterministic use of synthesis models to young clusters 
with masses larger than a few 10$^4$ M$_\odot$ in the optical domain \citep{CL04}, 
which corresponds to a limiting magnitude of $M_V = -11$ . In fact it 
is just a common-sense requirement: as an example, \cite{ZF99} reject 
point-like sources fainter than $M_V = -9$ mag in the analysis of 
clusters in Antennae, since it is the luminosity of single luminous 
variable stars. However, based on the LLL requirement, not only
 ``point-like" sources but {\it all} sources fainter than  $M_V = -11$ 
should be rejected in their analysis. For example, \cite{Pesetal08} 
show that young (200 Ma $<  t  < $1 Ga) 
clusters in the LMC do not fulfill  the LLL requirements, and therefore
the use of 
synthesis models within a deterministic framework is useless because
it may yield wrong results.

\subsection{Implications for the initial mass cluster distribution estimation}

In a recent pedagogical paper, \cite{Fall06} gives the relations between luminosity, 
mass and age distributions of young stellar clusters. A power-law 
luminosity function for young clusters is directly 
related to a power-law initial cluster mass function,  
 under the assumption that the mass-luminosity ratio depends 
{\it only} of the age of the cluster.
We have shown that this assumption is only valid for the case of massive 
clusters. The current controversy on the shape of the initial
mass function of clusters, where small differences between a possible
power-law or a log-normal distribution are important, depend
crucially on  clusters with masses around 10$^4$ M$_\odot$. 
As we have shown, this mass range is below the limit of 
application of synthesis models in a deterministic way, and a probabilistic 
framework is required for proper results, even though 
 it implies an intrinsic loss of precision.

{\it Acknowledgements}. We want to thank  Valentina Luridiana for comments and 
discussions and Mark Taylor for the development of TOPCAT, which has 
been used to examine and explore the results of the Monte Carlo simulations.
This work was supported by the Spanish project PNAYA2004-02703. 
MC is supported by a {\it Ram\'on y Cajal} fellowship and a visiting astronomer
position at GEPI, Observatoire de Paris.

\end{document}